\documentclass[10pt,a4paper,twoside]{article}
\usepackage{latexsym,amssymb,enumitem}
\usepackage[utf8]{inputenc}
\usepackage[english]{babel}
\usepackage{graphicx}
\usepackage{array}
\usepackage{tabularx}

\usepackage[T1]{fontenc}
\usepackage[intlimits]{amsmath}
\usepackage{graphicx,url,amssymb,amsthm,titlesec}
\usepackage{float}
\usepackage{placeins}

\begin{document}
\begin{center}{\Large\bf
Some optimization possibilities in data plane programming
}\end{center}
\begin{center}{\large\bf\noindent
Altangerel Gereltsetseg, Tejfel Máté
}\\[2mm]
3in Research Group, Martonvásár, Eötvös Loránd University, Budapest
\\[1mm]\texttt{
gereltsetseg@inf.elte.hu, matej@inf.elte.hu
}\end{center}

\vspace*{7mm}
\textbf{Abstract.}
Software-defined networking (SDN) technology aims to create a highly flexible network by decoupling control plane and the data plane and programming them independently. There has been a lot of research on improving and optimizing the control plane, and data plane programming is a relatively new concept, so study on it is one of the hot topics for researchers. At the 2019 Dagstuhl Seminar, well-known scientists on computer networking discussed challenges and problems in the field of data plane programming that need to be addressed over the next 10 years.  Based on this seminar issues and papers review, we suggested some possible solutions which are for optimizing data plane to improve packet processing performance and link utilization. The suggestions include (i) enriching data plane language with asynchronous external function, (ii) compression based on payload size, (iii) in-network caching for fast packet processing, and (iv) offloading external functions to an additional thread, virtual machine (VM) or server, etc.  In addition, we implemented some of these in the P4 data plane language to illustrate the practicality. 

\vspace*{1mm}

\textbf{Keywords. }Software data plane, data plane optimization, asynchronous packet processing

\section{Introduction}

The network data plane specifies the forwarding behaviour of the network device, and the control plane sets the rules for transmitting data in the network.  In a traditional network architecture, they are combined into a single device and have fixed-configuration because they are programmed only by device vendors. The programmable data plane in the SDN architecture has the following advantages over traditional fixed-configuration data plane.

\begin{itemize}[topsep=0pt,itemsep=-1ex,partopsep=1ex,parsep=1ex]
    
	\item \textbf{Flexible software development - rapid design cycle, fast innovation, fix data plane bugs in the field:} The development of a new protocol and network feature takes a long time in a traditional network because it passes through a multi-step process, such as a standard organization, a chip designer, and a software developer. For example, the introduction of the Virtual extensible LAN (VXLAN) protocol took four years from the time it was first proposed until it went into production [1].  In SDN, programmers can develop data plane applications faster and enrich them with the protocols and network features they design, regardless of the hardware manufacturer or standard organization.
    \item \textbf{Reducing complexity:} Unnecessary protocols, functions, and tables can be removed from a programmable switch, but a fixed configuration switch integrates everything, even if it is not involved in packet processing.
    \item \textbf{Modularity:} Forwarding behaviour (data plane) on software switch can be extended through external functions and libraries. 
	\item \textbf{Portability:} Specifying forwarding behaviour once and compile them to many devices [2]. 
\end{itemize}

The concept of data plane programming makes the network very flexible, but programmable switches are still not widely used in production networks for some reasons such as performance. Therefore, more optimization works that can improve network performance, security, and scalability on the data plane are needed to make it more practical. As mentioned above, A Dagstuhl seminar raised several questions on the data plane that need to be addressed in the next 10 years. These questions are improving performance and power efficiency, what type of processing is done in networks? , the right hardware architecture for programmability, what type of computation we add to the network, and where? ,  deploying and optimizing stateful network functions in software switches, a technique for improving software switch memory, the study of in-network computing, and improving network characteristics [3].    
There are also some related data plane programming researches that was currently conducted. We have categorized them into the related work section.

Our study suggested four ideas that can optimize plane data applications, and we created prototype implementations for two of them. The first idea is about how to integrate asynchronous compression functionality into data plane language, the second one is related to improving compression efficiency, and the last two are related to improving the performance of asynchronous packet processing. 

The rest of this paper is organized as follows. Section 2 discusses the related work of improvement and optimization of data plane programming. Section 3 gives a brief introduction about data plane programming and P4, asynchronous programming, and presents the motivations for optimizing data plane programming. Section 4 illustrates our optimization ideas in data plane programming and case studies showing their practicality. Section 5 presents conclusions and future works.

\vspace*{1mm}

\section{Related work}
Domain-specific programming languages with simple syntaxes and semantics have recently emerged for describing packet processing behavior (data plane) of programmable forwarding elements. Researches in data plane programming are often around these languages, and we generally classified as follows.
\begin{enumerate}[topsep=0pt,itemsep=-1ex,partopsep=1ex,parsep=1ex]
   \item The data plane programming languages (P4, Openflow, Protocol-Oblivious Forwarding (POF),  Domino)  themselves are still under development, so they need improvement and optimization in terms of syntax, semantics, language construct, Application Programming Interfaces (APIs) and how to optimally extend the language with external functions[3]. The latter case is one of the tasks we intend to implement, so we would like to cite two examples: The P4 language can be expanded in two ways through external functions. ROHC[4]  tested these two methods, and the Netronome Company expanded the P4 pipeline of Network Flow Processor (NFP) with a C function by developing a P4 backend compiler toolset [5], but both did not specify how to integrate asynchronous functions. Examples of language construct is pcube [6], which provides primitives to synchronize state variables across switches in distributed data plane applications.
   \item Because the network provides a wide range of services, it is challenging to create a single optimized, programmable switch or super network solution. Therefore, there were some experiments to create application-based software switches that are optimal for a particular service. DC.p4 [7]  explored existing P4 constructs and suggested additional language constructs that can be added to the P4 language for building an optimal data center switch.
   \item However, it is also important to look for opportunities to develop general optimization methods that can be addressed in data plane programming to improve network or packet processing performance, flexibility, scalability, security, link utilization, and reliability. For example, A zero flow [8] uses an idle timeout to remove corresponding flow entries from the flow table to make it scalable, Dargahi et al [9] made a survey on the security implications that data plane programmability brings about, and policy-driven optimization [10] is recommended to activate only the required functions on the P4 switch to increase the efficiency of the pipeline. 
   \item There are researches on developing new protocols and features or optimizing the old protocols and data plane functions, and determining where to implement them more effectively, for example, whether to implement  on a network device itself, on an external device, or at which level of TCP / IP protocol suite. For instance, Paxos protocol for a distributed network, which was previously implemented in the application layer, was tested at the network layer (software switch) [11]. Moreover, the data plane algorithms, such as congestion control, scheduling, and traffic engineering, are often built into the hardware to achieve line speeds but can now be implemented in the software with the help of data plane programming. In addition, there are some attemps to deploy stateful packet processing in software switch. TEA[12] suggests architecture that enables state-intensive network functions on programmable switches, and  SNAP[13] offers a simpler “centralized” stateful programming model allowing programmers can implement a broad range of applications, from stateful firewalls to fine-grained traffic monitoring. Therefore, in addition to developing new things, software developers have the opportunity to test and optimize implementations, which are already strictly standardized in the traditional network, in other ways.
   \item Building software switch with  high performance packet processing by using some techniques, and external libraries for fast packet processing such as Data Plane Development Kit (DPDK), and netmap: VPP[14] is framework implemented in user-space with kernel- bypass mode, offering flexibility of a modular router the benefits brought by techniques such as batch processing that have become commonplace in lower-level building blocks of high- speed networking stacks (such as netmap or DPDK). As a result, it gave the reasonable packet forwarding performance. T4P4S[15] design with DPDK library gave possibility to extend software switch with high speed functionalities. 
   \item Optimizing data plane applications with Network functions virtualization (NFV). It can improve the flexibility of network service provisioning and reduce the time to market of new services [16].  In [17], NF instances can be chained on demand, directly on the data plane by extending SDN to support stateful flow handling based on higher layers in the packet beyond layers.  A detailed evaluation of this implementation shows that popular NFs are working more efficiently than other alternative solutions.  
   \item Monitoring and auditing on data plane and its programs: With the advent of programmable data planes, the In-band Network Telemetry network monitoring framework, is originated on the data plane without the involvement of a control plane. Because it is performed directly on the data plane, the performance measurements are more accurate and real-time compared with traditional solutions[18].  Data plane programmability can bring the risk of introducing hard-to-catch bugs at a level that was previously covered by well-tested devices with a fixed set of capabilities. p4pktgen[19] is a tool for automatically generating test cases for P4 programs using symbolic execution. These test cases can be used to validate that P4 programs act as intended on a device. Performance guarantees for P4 through cost analysis[20] classify, and evaluate implementation-level cost models for P4 programs, enabling developers to estimate program execution costs (with CPU cycle precision) before actually deploying the solution to hardware.

\end{enumerate}
\vspace*{1mm}

\section{Background and motivation}
\subsection{Data plane programmability and P4}
P4 is a domain-specific language developed for data plane programmability. It lets us control switches "top-down" by first specifying their forwarding behavior, then populating the tables we've defined.  Besides, the P4 compiler automatically generates the APIs needed to populate the table. The programmable building blocks in this language can be a parser, a match-action tables, a deparser, and so on.  The parser block specifies the header sequence. The specified headers are examined in match-action tables and based on an examination, appropriate actions are executed on the packet, such as dropped, modified, or forwarded. For example, if the packet is determined to be forwarded in the match-action phase, then the outgoing packet is constructed in the deparser block. These types of construction blocks are programmed in P4 language flexibly and simply [21].  

The main advantages of the P4 are protocol and target-independence. The protocol-independence means that the programmers can customize network protocol headers and extend this language by adding external data plane functions.  P4(16) includes the extern primitive, which provides an interface to access external libraries and functions, such as checksum computation and cryptographic operations.  External functions are a very powerful concept but are not fully specified in P4, for example, how to deploy network functions in an asynchronous way. 

In terms of a target-independent feature, P4 code can be compiled on many kinds of P4 targets (devices) such as general-purpose CPUs and other kinds of network devices. To run a P4 program on a new target, the developers can customize the back-end compiler of P4 or add a new converter after the backend compiler (Figure 1). In other words, each target must be provided along with a compiler that maps the P4 source code into a target. The target-independence is implemented in this way.

\graphicspath{ {./images/} }
\begin{figure}[h]
\centering
    \includegraphics{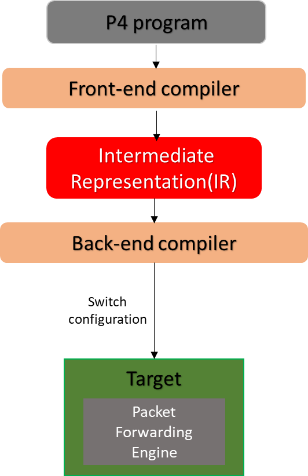}
\caption{P4 workflow }
    \label{Fig:}   
\end{figure}

The P4 allows great flexibility in the description of packet structure and processing, but it still needs to be improved to optimize packet processing and increase network performance.

\subsection { Network characteristics and some software optimization techniques } 
The packet forwarding rate, packet drop rate, packet processing delay, link utilization, jitter, and so on are the metrics used to evaluate network performance. By maximizing or minimizing some of them, we will create a more optimal network with higher performance. These characteristics depend on network devices, software, hardware and network architecture, and network organization. In other words, it is necessary to develop optimal solutions at the software and hardware level to improve these characteristics.  Software switches offer flexibility to service providers but potentially suffer from low performance.  Data plane programming lets the developers introduce new features and protocols into the network without waiting for new hardware or vendor-specific solutions. Therefore, software developers can create optimal solutions to improve the network characteristics in various ways at the software level.

Some performance enhancement techniques in the network are quality of services (QoS), load balancing, caching engines, and offloading computation to an external server, compression, and asynchronous programming. In the next section, let me briefly introduce asynchronous programming because we used it in our solution.

\subsection { Asynchronous programming }
Developing synchronous programming is simpler than asynchronous programming. Asynchronous programming is more complex to design and implement, but in some cases, it has higher performance than synchronous programming. Therefore, it is important to balance simplicity and performance when developing software. Nowadays, the coroutine and future/promise asynchronous programming techniques are used to build database and web server programs. These solutions manage the complexity of asynchronous programming while achieving good performance [22].	

Network functions can be classified as compute-intensive, state intensive, non-compute intensive, or stateless. Based on these properties, the network functions can be performed by the network device itself (in-network computing), by an external device, asynchronously, or synchronously. For example, firewall, Network Address Translation(NAT), Intrusion Detection System[IDS], and proxies often need to contact external services while processing network flows, e.g, querying external databases or saving critical per-flow states on reliable storage to resist failures. To ensure high speed packet processing while executing external queries, these NFs must fully exploit asynchronous programming: after generating request to external service, the NF shouldn’t block-and-wait for the response in a synchronous fashion: instead it should save the current processing context, register a callback function to handle the response upon its return, and switch to process other task. 

Also, firewall and load balancing functionalities are considered state and compute-intensive functions because they  require more processing in the system.  Therefore, they can be implemented using asynchronous programs to run as a separate processing component on the network device itself(if it has high processing capacity), or on a separate serving instance. 

In addition, the current network processors can support aggregate throughput of tens of gigabits per second, but they are mostly optimized for processing packet headers and leaving packet payloads untouched. Therefore, it is important to look for ways to optimally organize the payload processing functions such as encryption and compression on the data plane to achieve high performance. Data plane programmability is a good experimental ground that allows network functionality to be run in a variety of ways.

\section {Some optimization possibilities in data plane programming}
Our study aimed to suggest some general methods that can improve packet processing performance and link utilization by optimizing data plane applications. In the following section, we presented the proposed methods and the possible implementations for some of them.

\subsection{ Optimization 1: Enriching data plane language with asynchronous external function }
As mentioned in the previous section, network functions process packets based on header information or payload. Functions based on payload processing require higher performance. If these functions are not run asynchronously, performance efficiency may be limited. Most data plane languages do not specify how to integrate an asynchronous external function into a packet processing pipeline. P4 standardizes the use of an extern construct to add an external function but does not support asynchronous execution.  Therefore, there is a need to develop an effective, simple, and general way to add asynchronous external functions to the data plane language. We proposed a possible solution to this issue and shown in a case study in which the asynchronous compression function was integrated into the P4 language using the coroutine technique. We described the case study in detail in Section 4.5.1.

We proposed a possible solution to this issue and shown in a case study in which the asynchronous compression function was integrated into the P4 language using the coroutine technique. We described the case study in detail in Section 4.5.1.

\subsection{Optimization 2: Compression based on certain criteria}
We implemented optimization 1 as an example of an asynchronous compression function. Packets are compressed to improve the link utilization, but the compression of all packets may increase the computational load of the device, so criteria can be set as to which packet is best to compress. Based on the relevant research, we have determined that the following criteria can be set.

\begin{itemize} [topsep=0pt,itemsep=-1ex,partopsep=1ex,parsep=1ex]
   \item[(a)] Packets with a larger payload (more than 500 B) produce a better compression ratio[23] [24]. 
   \item[(b)] If an intermediate node receives a packet with a larger payload, it can calculate the waiting time of the packet in the queue. If the waiting time is greater than a certain instantaneously calculated threshold, the node will compress the packet’s payload [25].  
\end{itemize} 
The implementation of this idea is described in Section 4.5.2.

\subsection{Optimization 3: In-network caching}
The cache is used to make packet processing faster because it skips some look-up process of the packet on the Central Processing Unit (CPU). Furthermore, in-network (network device itself) caching functionality has been shown to run at line rate on programmable switches [26].  Our case study (optimization 1) can apply the cache as well, and section 4.5.3 explains in more detail how cache can increase the performance of the asynchronous operation.
\subsection{Optimization 4: Offloading external functions to an additional thread, VM or server }
The Network Function Virtualization (NFV) relocate network functions from dedicated appliances to generic servers or virtual machines to speed up the deployment of new network services. Besides, offloading some of the computational tasks of network devices on external devices can help to reduce the load on the core device. The NFV can be implemented without SDN but it is more valuable if these two concepts and solutions can be combined. As mentioned above, the P4 language itself assumes that some actions or functions are better done using externs. The extern can be treated as black boxes, implemented using accelerators or special-purpose external servers. This is somewhat similar to the concept of NFV. There are some attempts in this direction. For example, Netronome company deployed P4 based applications in Server-Based networks. They aimed to create a solution similar to a package management system on Linux. The network data plane functions are developed as a package and distributed through the repository. They should be installed on the server or external device, and the installed function should have cooperated with the core packet processing applications [27].

\subsection{Implementation}
\subsubsection{Extending P4 pipeline through asynchronous compression function (Optimization 1)}
We made a prototype implementation of optimization 1 on DPDK based software switch. This switch was built using the T4P4S converter for P4 which was developed at ELTE [15]. The T4P4S architecture allows the use of external libraries, and DPDK libraries are used for packet input and output of this switch, so we can say that it  is a software switch with high performance. T4P4S architecture (Figure 2) has two main abstraction layers, and external functionalities are implemented in the hardware-dependent abstraction layer. Therefore, we integrated external compression functionality into this layer.

\graphicspath{ {./images/} }
\begin{figure}[h]
\centering
    \includegraphics{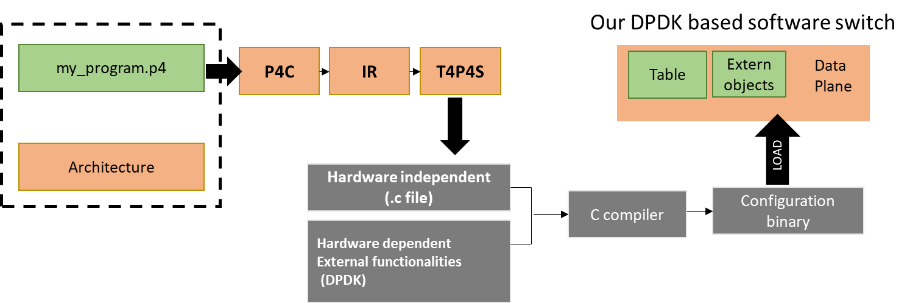}
\caption{Our DPDK based software switch}
    \label{Fig:X1}   
\end{figure}

Now let me explain how we created a P4 application with external asynchronous functions (compression) on this switch and how it works. Our application has two main loops: main packet processing loop with P4 pipeline, and compression/decompression loop, in short, external function loop. In the following sections, we described these two loops and how these two loops interact asynchronously.

\textbf{Main packet processing loop with p4 pipeline:} 
Our P4 code has three main control parts: parser, ingress, and deparser. We have added external functions into the P4 architecture (v1model)  by using the extern construct: extern void $async\rule{1mm}{0.15mm}comp()$ and extern void $async\rule{1mm}{0.15mm}decomp()$. As mentioned in the previous section, we have defined these functions in the hardware dependent layer of T4P4S. The following picture shows our P4 pipeline with external functions. Our switch can process packets with or without calling external compression based on layer 2 address and newly defined tag field.  In other words, the P4 pipeline will check the tag field to decide whether or not to compress the packet.

\begin{center}
\graphicspath{ {./images/} }
\begin{figure}[h]
\centering
    \includegraphics{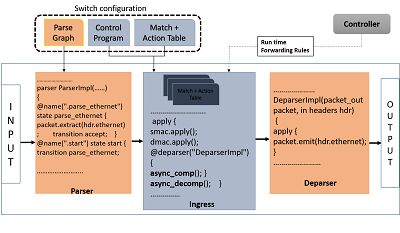}
\caption{Packet forwarding abstract model in P4 based software switch}
\end{figure}
\FloatBarrier 
\end{center}

We can see that  $async\rule{1mm}{0.15mm}comp()$ and $async\rule{1mm}{0.15mm}decomp()$ asynchronous functions are invoked at the ingress part from figure 3.

\textbf {External compression function loop:} As mentioned above, an external compression function was integrated on a hardware-dependent abstract layer of the T4P4S compiler. Our research team also integrated asynchronous encryption on this software switch[28]. Table 1 shows a comparison of compression and encryption implementations. DPDK ZLIB Poll Mode Driver (PMD) was run as a compression function, and DPDK OpenSSL PMD was run as an encryption function.

\begin{center}
\begin{table}[ht!]
\begin{tabular}{|l|l|}
\hline
\textbf{Compression function}                                                                                                                                                                                                             & \textbf{Encryption function}                                                                                                                                                                 \\ \hline
\multicolumn{2}{|l|}{Creating virtual devices with corresponding memories by using DPDK  libraries.}                                                                                                                                                                                                                                                                                                                                     \\ \hline
\multicolumn{2}{|l|}{Loading compression/encryption PMDs on virtual devices.}                                                                                                                                                                                                                                                                                                                                                            \\ \hline
\begin{tabular}[c]{@{}l@{}}Configuring devices such as device id, \\ queue pair, defining a compression scheme\\  with parameters (compression algorithm, \\ compression type (stateless or stateful),\\  checksum and etc.)\end{tabular} & \begin{tabular}[c]{@{}l@{}}Configuring devices such as device id, \\ queue pair, defining encryption session with \\ parameters (encryption algorithm, cipher \\ key, and etc.)\end{tabular} \\ \hline
\multicolumn{2}{|l|}{\begin{tabular}[c]{@{}l@{}}Compression and encryption operations are represented by structures \\ which include all parameter for the intended operations.\end{tabular}}                                                                                                                                                                                                                                            \\ \hline
\begin{tabular}[c]{@{}l@{}}Defining compression operation such as \\ source, destination buffer, and offset. \\ The offset  determines from which byte \\ to start compressing.\end{tabular}                                              & \begin{tabular}[c]{@{}l@{}}Defining encryption operation such as \\ source buffer and offset. (Destination buffer is \\ unnecessary in encryption case)\end{tabular}                         \\ \hline
\begin{tabular}[c]{@{}l@{}}Attaching compression parameters to \\ compression operation.\end{tabular}                                                                                                                                     & \begin{tabular}[c]{@{}l@{}}Attaching encryption operation \\ to  encryption session.\end{tabular}                                                                                            \\ \hline
\multicolumn{2}{|l|}{Loading/enqueuing packet with defined compression/encryption operation on a virtual device.}                                                                                                                                                                                                                                                                                                                        \\ \hline
\multicolumn{2}{|l|}{\begin{tabular}[c]{@{}l@{}}The compression/encryption operations are performed when the enqueue API is invoked \\ on virtual devices.  A compressed/encrypted (processed) packet is placed in the\\  rte\_ring buffer of the virtual device.\end{tabular}}                                                                                                                                                          \\ \hline
\multicolumn{2}{|l|}{\begin{tabular}[c]{@{}l@{}}Retrieving/Dequeuing processed packets through  dequeue burst API on the virtual device \\ and redirected it to the main packet processing loop and this loop forwards the processed \\ packet to the exit interface.\end{tabular}}                                                                                                                                                      \\ \hline
\end{tabular}
\caption{Comparison of compression and encryption implementations}
\label{table:1}
\end{table}
\FloatBarrier 
\end{center}

As shown in Table 1, the general activity of the encryption/compression PMD is similar, with the difference being the compression and encryption operations: the encryption is session-based, and the compression is buffer-based. They were both implemented in a stateless manner.
\vspace*{5mm}

\textbf {4.5.1.1 Bringing asynchronous network function to the P4 data plane \vspace*{3mm}} 

In our case study, we used a coroutine paradigm that is able to execute the synchronous programs in a fully asynchronous fashion. As mentioned above, our code was two loops, the main packet processing loop and external function. Asynchronous packet processing is done as follows. Whenever a packet that should be compressed comes into our switch, the P4 pipeline processes this packet, and the main packet processing loop saves packet context information. Then, this packet is sent to the external function loop. The external function loop compresses the packet and sent it back to the main packet processing loop. The main packet processing loop resumes context information, and it sends the compressed packet to the exit interface based on the context information. In our solution, we have redirected it into a decompression switch. The running way of the decompression switch is the same as the compression switch. Figure 4 describes these operations.  
\begin{center}
\graphicspath{ {./images/} }
\begin{figure}[h]
\centering
    \includegraphics{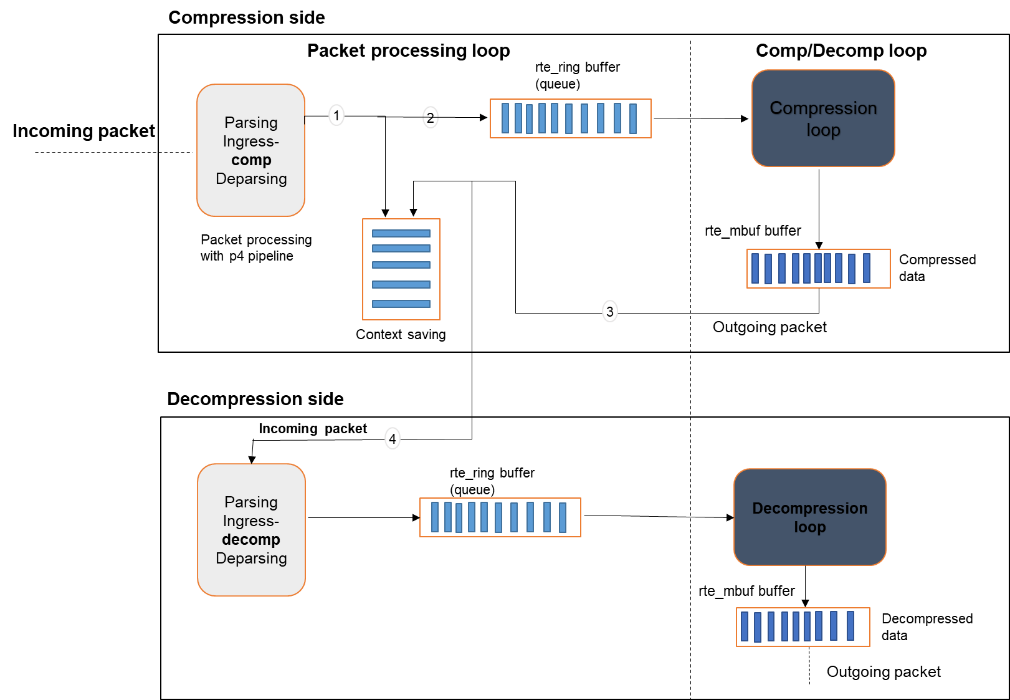}
\caption{Packet processing in our software switch}
    \label{Fig:X3}   
\end{figure}
\FloatBarrier
\end{center}
While external function is executed, other packets are transmitted normally through the main packet processing loop. We used a standard C library(the user flow context) to enable the coroutine operation. This library integration with DPDK made a feasible ground for building asynchronous packet processing. That integration is then embedded into the P4 data plane with the help of a T4P4S converter. This solution can be broadened with other functionalities such as anomaly detection, and network address translation (NAT).  We also offered some optimization possibilities in our solutions to increase efficiency and performance in the below sections. 
\subsubsection{Implementing compression based on payload size (Optimization 2) }
We extended the previous implementation to test this idea. First, we declared the threshold value (500bytes) of the payload to compress the packet, and second, we added a new field named payload length to the IP header in the P4 code because P4 allows us to define different kinds of fields in protocol. But we can also use standard total length field including payload and header in IP protocol. Then we compared these values at the ingress of the pipeline right before calling the external compression function, and if the length value is greater than the threshold, the asynchronous compression function is called, otherwise, the packet is processed normally and passed to the output interface.

\begin{center}
\graphicspath{ {./images/} }
\begin{figure}[H]
\centering
    \includegraphics{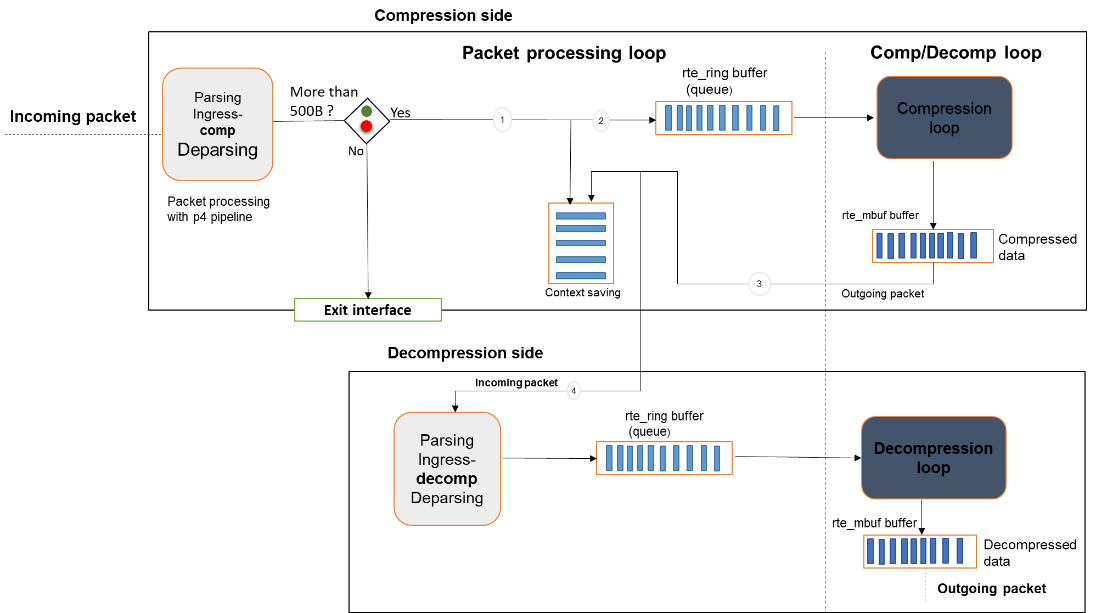}
    \caption{ Compression based on payload size}
    \label{Fig:X4}   
\end{figure}
\FloatBarrier
\end{center}


Figure 5 shows the scenario. When the packet comes to the switch, it is checked by the P4 pipeline, and if the payload is greater than 500B, it is redirected to the compression loop, but the context information must be stored before that.
\subsubsection {Possible implementation on in-network caching (optimization 3) }
Our case study (optimization 1) processes packets as follows: when a packet comes into a software switch, it is processed through the main loop (for example, the main loop defines exit interface), then this loop saves the context information of the packet, and it sends into compression loop. Once the packet is compressed, it will return to the main loop, resume the context, and pass the packet to the defined interface. If the context was cached in memory to be reused to process the next packet of one flow (the group of the packet with the same source and destination address), the packet processing can be faster.  Every packet is inspected by CPU.  

We should consider two issues to implement this idea: how to cache context information and how to distinguish one flow from another. We can program that context information of the packet will be cached for a certain period of time. If it is used again within the specified period, the period can be extended. If not, it can be dynamically deleted from memory when the time runs out.

Suppose there was one flow with 10 packets. The first packet is fully checked by the P4 pipeline, after which the context is saved. When the second and subsequent packets arrive, bypass the look-up process and pass it to the compression loop without saving the context. When the compressed packet returns, it is processed using the cached context. The context contains the routing information needed to process the packet, for example, the egress interface, next-hop IP address, and so on. In general, it looks similar to a routing table but the lookup structure is more efficient, so the packet forwarding is faster. In addition, the CPU is not involved  in packet transmission using the cache. However, to do this, it is important to define the packets as a single flow, and perhaps an additional label can be added to the packet to distinguish it. We envisioned an implementation in this way, and although it is not fully implemented, it seems straightforward. 

We envisioned an implementation in this way, and although it is not fully implemented, it seems straightforward.

\subsubsection { Possible implementation on offloading external functions to external device (optimization 4)}
The packet I/O is done on a master thread, and the main packet processing loop and compression function are running on the slave thread in our case study.  In the future, to run an external function that requires a lot of computation, we can run that external function on an additional thread or some external devices.  According to the DPDK document, it is possible to load each function in a different thread, but this is not intended to optimize performance and is just demonstrated the practicability of doing this.  We summarize the proposed solution in the following figure. If the switch operation is expanded or the switch load is increased, the switch resource may be expanded by an external device as shown in figure 6, and for example, two compression loops are running on the external devices. If the load on the first server increases or the first server has a problem,  it can be configured to move to the second server. In this way, other functions can be distributed on an external device.

\begin{center}
\graphicspath{ {./images/} }
\begin{figure}[ht]
\centering
    \includegraphics{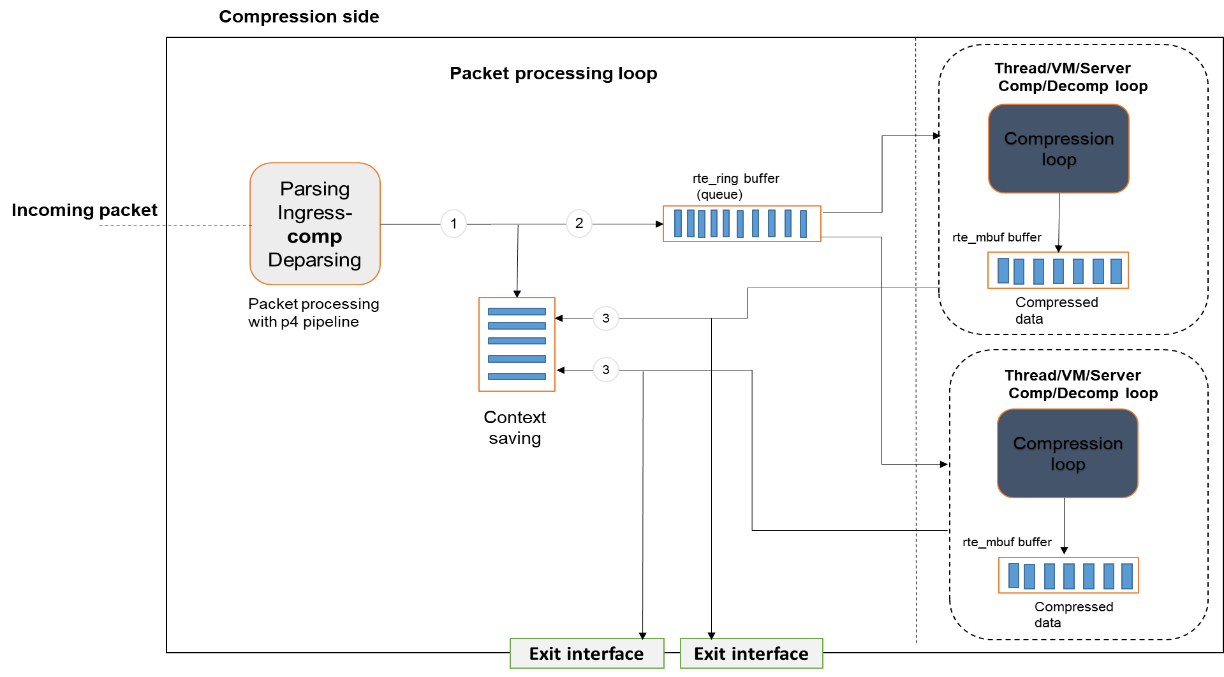}
\caption{Figure 6. Offloading network function to external device}
\end{figure}
\end{center}

\section {Conclusions and future work }
Data plane programming is one of the explosions in computer networking, but it is not fully matured and is evolving through different kinds of research and experiments. We presented some optimization possibilities on data plane programming and created some prototype solutions for these ideas in P4. These ideas are interrelated, for example, our research group first created a high-speed packet I/O software switch with a P4 pipeline by using commodity hardware, then we expanded the P4 pipeline through an asynchronous compression functionality based on the previous implementation of the encryption case, and then, to increase the efficiency of compression, we suggested criteria to compress packet, and the next two ideas are about improving the performance of asynchronous packet processing.  

Our prototype solutions are stable, modular, and well-defined, so they can be partially changed or improved in the future. For example, we can change asynchronous programming techniques or compression algorithms without completely modifying the core code. Also, we have used a generic asynchronous method applicable for encryption, compression, and other functionalities. 

Although we have not made a performance analysis, in general, synchronous performance exceeds asynchronous performance when the computational cost of an external function is low. Compression and encryption functions do not require high computational costs compared to network anomaly detection, proxy and firewall functionalities.  However, we believe that the practical implementation of this idea in P4 is a big step forward. Therefore, we can add other computation-intensive external functions, such as anomaly detection in the pipeline to the data plane, in the same way as adding compression.

In the future, we will extend/validate this work with our current study: Detection of Internet of Things (IoT) attacks using machine learning approaches based on In-Band Network Telemetry (INT) data. INT is a novel monitoring application developed using a programmable data plane, that can collect network characteristics (INT data) in real-time without affecting network performance. As far as we know, this is the first attempt to use INT data to detect IoT anomalies. We consider it is possible to implement it asynchronously in the data plane just like adding a compression function.

After implementing this, we can compare the performance results with the compression case. Similarly, we believe that there should be many applications of using an asynchronous functionality in the data plane.

\setlength{\parindent}{4em}
\paragraph {Acknowledgement} The research has been supported by the European Union, co-financed by the European Social Fund (EFOP-3.6.2-16-2017-00013, Thematic Fundamental Research Collaborations Grounding Innovation in Informatics and Infocommunications).

\end{document}